# The Quest for Ultimate Broadband High Power Microwaves

Andrew S. Podgorski, *Fellow, IEEE*

*Abstract*—This paper describes High Power Microwave (HPM) research into combining GW peak power to achieve MV/m and GV/m radiated EM-fields that surpass the previously establish limits of air and vacuum breakdown fields in the 1 to 500 GHz frequency band. To achieve such fields multiple independently triggered broadband GW power sources, supplying power to multiple spatially distributed broadband radiators/antennas are used. The time of triggering of each generator and generated pulses spectral content are chosen to enable varying of the radiated EM-field and energy to achieve the most efficient level of electromagnetic interaction at the point of power delivery. A single TW antenna array generating a MV/m EM-field is used as an ultimate microwave weapon in the 1 to 5 GHz range ensuring the highest probability of target destruction at 10's of kilometers. Use of multiple TW antenna arrays allows the scaling of power deliver into the PW power range and, if properly focus, will provide a GV/m radiating EM-field at resonance plasma frequencies in 300 GHz range creating a new paradigm for molecular and atomic research leading to fusion power.

## I. INTRODUCTION

Introduced in 1992, the Composite Threat (CT) concept [1, 2] identified the limit of ultimate electromagnetic (EM) threat due to air and vacuum breakdown. To reach tangible (Ultimate) threat further research in the area of Terawatt broadband short pulse EM was needed. Although at the time no Ultimate threat capabilities existed, testing methods and equipment for protecting against the CT were developed [3, 4]. Further development of narrowband and broadband high power generators led to achieving GW peak power at microwave frequencies [5]. However, use of narrowband GW HPM systems with inefficient delivery of the EM energy to the target undermined the efficacy of deploying the HPM for military applications [6]. Until today, broadband HPM weapons use a single GW generator to increase the coupling efficiency but fail to acknowledge the need for TW power. This failure to consider TW power undermined the goal of achieving significant improvement in the effectiveness of HPM weapons.

Currently the maximum HPM EM-field levels defined by Military standard MIL-STD-464C (Dec. 1, 2010) [7] shown in Fig. 19 of an Appendix, indicate that the most severe narrowband intra-system EM environments are limited to EM-field peak levels of 28 kV/m$_{RMS}$ in the 2.7 to 3.6 GHz frequency range. The same standard states that for intentional threats at a distance of 1 km and frequencies from 8.5 to 11 GHz, the narrowband EM-field peak level is limited to 69 kV/m. However, due to increasing concern regarding the values specified in the standard, the MIL-STD-464C cautions that the narrowband intentional threat specifications "should be verified prior to implementation." Considering the above concerns and uncertainty in recognizing the magnitude of future possible threat, this paper follows the CT [1, 2] stipulating that if TW peak power limited by air or vacuum breakdown field is reached the HPM threat could be increased by few orders of magnitude (See Fig. 1). Initial experiments using a single beam [8] demonstrated that focusing 1 GW, 100 ps pulse into a 3 cm diameter focal point resulted in achieving a 10 MV/m field corresponding to previously identified air breakdown limit of 30 MV/m. These experiments provided a proof of concept that merging GW power from individual generators using broadband antenna arrays [4] enables the HPM system to illuminate larger targets with Ultimate Threat TW radiated power. For long distance illumination, broadband Cassegrain antenna arrays capable of generating and radiating TW peak power are used. By deploying multiple collimated beams from TW Cassegrain antennas, it is possible to achieve PW-radiated power exciting plasma at a fusion plasma frequency of 300 GHz.

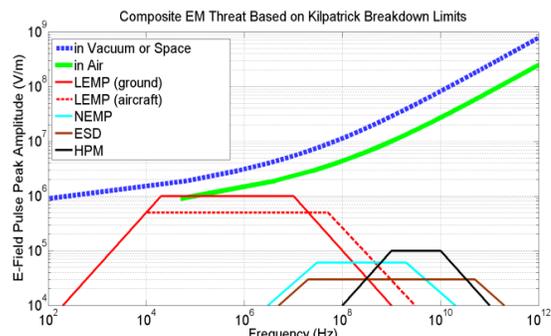

Fig. 1. Composite EM threat, based on breakdown E-field measured in microwave cavities, in comparison to HPM, LEMP, NEMP and ESD threats. The frequency is equivalent to the pulse duration, $f=1/T$.

In summary, Chapter II describes the achievable peak pow-er of broadband HPM generators limited by the breakdown of water coaxial capacitors used in pulse forming switches. Chapter III addresses broadband TEM-horn antennas, while Chapter IV addresses broadband HPM TEM-horn arrays, Cassegrain and Collimated Cassegrain arrays. Chapter V describes collimated broadband HPM spherical arrays for plasma studies and Chapter VI contains conclusions.

## II. BROADBAND HPM GENERATORS

Currently Marx generators using a water capacitor in the high voltage switch enable the HPM system to deliver the highest peak power and fastest pulse rise time at HPM frequencies. Use of water is dictated by its energy storage properties and the ability to modify switching rise and recovery time. This paper details experiments in which the water capacitor breakdown EM-field was established using the Sandia National Laboratories test of dielectric-breakdown of water insulated 6 MV, μs transmission line generators [9] and author's research on GW Marx generators [10, 11] (See Fig. 2). The estimated maximum peak power level based on the breakdown EM-field of the water capacitor (Fig. 3) accounts for the geometry of an optimally designed water capacitor and switch assembly.

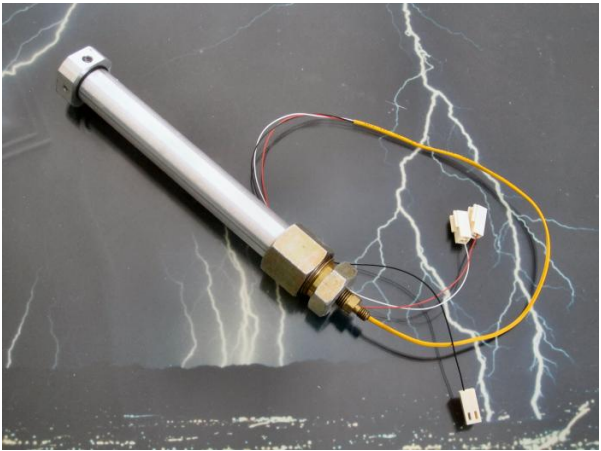

Fig. 2. View of 1 GW Marx generator, 10 ps rise time, 100 ps duration.

Fig. 3 shows EM-field breakdown of a coaxial water capacitor, being a part of a 100 ohm generator, as a function of frequency $f=1/T$ with $T$ as the pulse duration. The concurrence of the Sandia minimum accelerator limit (no-breakdown) in the MHz region with the author's achieved limits in ns and ps generators established the breakdown field. Based on Fig. 3 and design criteria for the best geometry of generators, the achievable peak power levels of pulse generators at 100 Ω are:
 - 1 to 5 GHz, 160 GW, 100 ps rise time, 1 ns duration,
 - 10 to 50 GHz, 6.4 GW, 10 ps rise time, 100 ps duration,
 - 100 to 500 GHz, 0.2 GW, 1 ps rise time, 10 ps duration.

For the maximum EM impact on target, the design of the Marx generator has to be optimized to achieve the highest peak power (*P*), longest pulse duration (*T*), and wide bandwidth. However, since the power factor $k_p=P*T$, and the bandwidth factor $k_\gamma=f_{max}/f_{min}=f_{max}*T$ are interrelated, a compromise between bandwidth and power has to be considered. To assure the broadband nature of generated pulses the minimum bandwidth factor should be $k_\gamma \geq 3$. As such, when estimating the maximum achievable peak power level of pulse generators the bandwidth factor $k_\gamma=5$ was used.

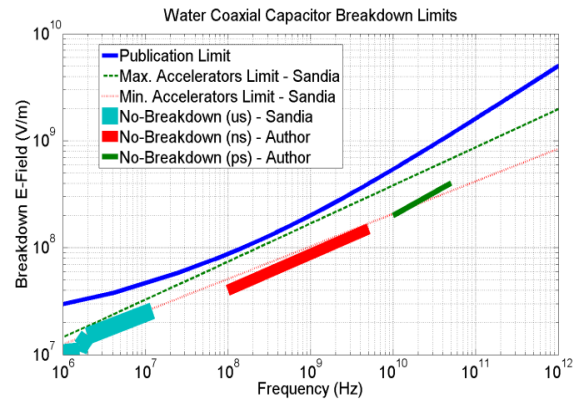

Fig. 3. View EM-field breakdown limits of coaxial water capacitors in function of frequency equivalent to the pulse duration - $f=1/T$.

## III. BROADBAND HPM TEM-HORN ANTENNAS

To achieve the highest impact on target the radiated EM-field has to be in the range of MV/m corresponding to air and vacuum breakdown levels. Such EM-fields are achievable by merging the GW power from individual generators using broadband antenna arrays formed from multiple broadband high power TEM-horns. Although many present broadband systems use TEM-mode antennas [12], there is a fundamental difference between existing broadband antenna systems and those presented in this paper. The TEM-horn developed for this research is a hybrid consisting of TEM-mode and a horn antenna and to achieve TW power the generator with pulse forming switch is separated from the antenna. All TEM-mode antennas radiate a conical wave front between the plates and the non-constrained external field. Placing the TEM-mode antenna inside the horn increases antenna efficiency from 30% to 50%, improving gain and increases field confinement reducing coupling interference between antennas within an array (See Fig. 4) [13, 14, 15, 16]. Thus, it is possible to build more effective EM arrays. To provide a uniform circular and spherical TEM wave front in the frequency band of greater than one decade the TEM-horn must be optimized in several ways including the cross-section, the length of the antenna enclosure and TEM-septum width and height.

In the TEM-horn, the radiated conical wave front for each frequency initiated at a different focal point progresses linearly from the mouth of the antenna to the back assuring an increasing frequency undistorted phase uniformity. As such, the length of the TEM-horn is proportional to the required bandwidth. EM absorbers installed at the mouth of the antenna horn confine the conical beam and through attenuation of the side-lobs make it viable to use optical-ray analysis for the design of individual horns and the entire array. Initialy the new TEM-horn antennas used in the NEMP, HPM and EMI/EMC test facilities (Fig 5 and 6) [17, 18], were optimized to achieve the ±3 dB E-field uniformity in the near-field, making them suitable for the design of complex antenna arrays. In order to deliver the highest EM-field on target, one must place as many TEM-horns in the array as possible by minimizing the cross-section of the mouth. Fig. 7 represents -3 dB radiated beam diameter for TEM-horn having different mouth widths in function of normalized frequencies indicating that the mouth width of 2.25 λ delivers the required bandwidth and uniform change of beam diameter.

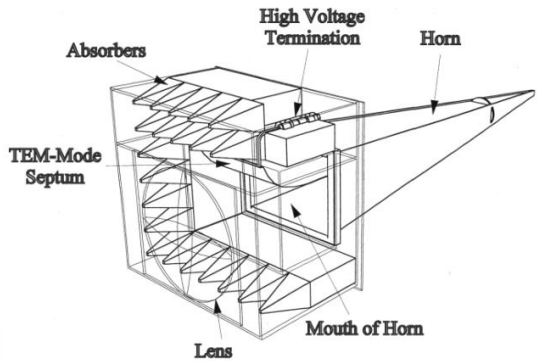

Fig. 4. Cutout view of the high power, Pulse and CW, TEM-horn.

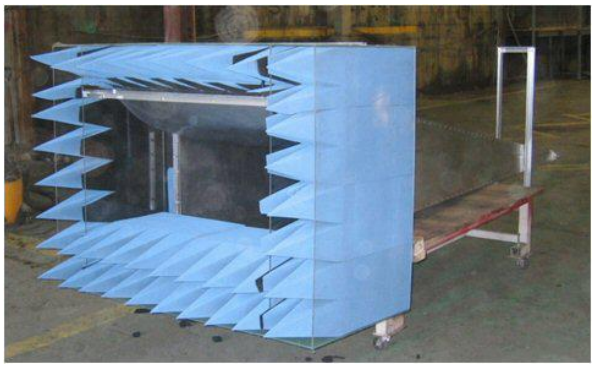

Fig. 5. TEM-horn: 5 GW, 100 MHz to 5 GHz.

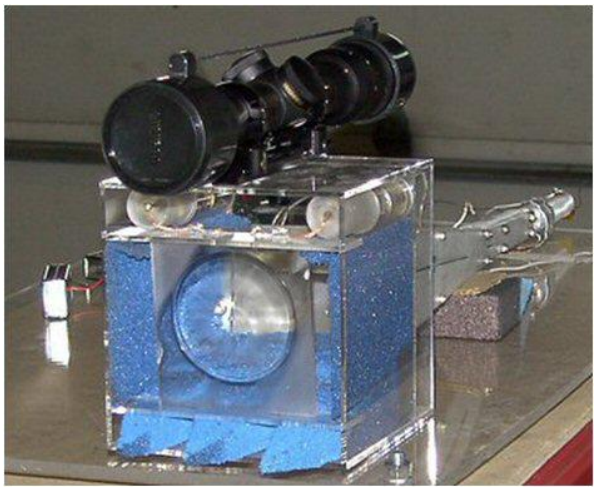

Fig. 6. TEM-horn: 1 GW, 5 to 50 GHz and externally triggered Marx generator powered from small 9 V battery.

Use of wavelength and frequency $f$ normalized to the central frequency $f_c$ permits optimization of the TEM-horn mouth width in the entire electromagnetic spectrum.

$$\frac{f}{f_c} = \frac{f}{\sqrt{f_{min} * f_{max}}} \qquad (1)$$

To assure constant beam divergence in function of frequency, the slope of -3 dB beam diameter at target expressed in wavelength should be close to two. However, such slope required for large mouth width limits the number of TEM-horns placed in an array - a compromise is required. According to Fig. 7, the mouth widths of 1.5 and 1.875 λ do not guarantee operation at low frequencies but the optimal width of 2.25 λ allows low frequency operation, placement of the highest number of antennas in the array and the smallest beam divergence at the target. Fig. 8 shows the antenna gain and the main beam angle in function of normalized frequency. At the central frequency ($f/f_c$=1), the gain of TEM-horn, having mouth width of 2.25 λ is approximately 17 dB and is comparable with the gain of a standard microwave horn of a similar size. As per Fig. 8, the beam angle decreases with increasing frequency of operation limiting the illuminated area at the target. Since immunity testing requires illumination of the entire target at the maximum frequency of operation, the beam diameter and the target diameter at the maximum frequency should be the same. However, since HPM weapons require largest EM-field over the entire target the beam diameter to target diameter ratio needs to be optimized.

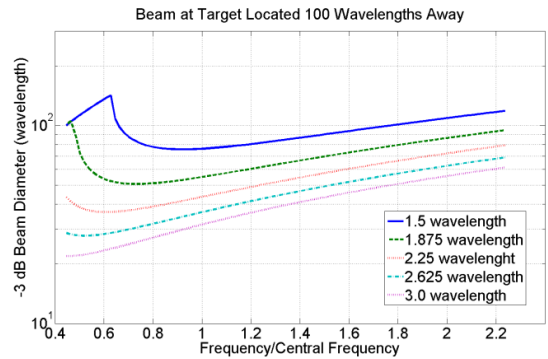

Fig. 7. TEM-horn -3 dB radiated beam diameter at 100 λ from the antenna in function of normalized frequency - calculated for different antenna mouth width defined in λ's at central frequency.

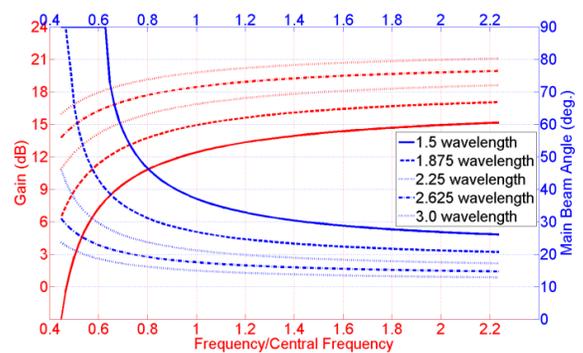

Fig. 8. TEM-horn gain and beam angle in function of normalized frequency - calculated for different antenna mouth width defined in λ's at central frequency.

Considering the peak power levels of the pulse generators seen in Chapter II, and assuming 50% antenna efficiency, the calculated peak E-field generated by single antenna for frequencies from 1 to 500 GHz at distance of 100 λ is presented in Fig. 9. As per Fig. 9, the peak E-field increases in function of frequency and at the central frequency; in 1 to 5 GHz - 1 MV/m, in 10 to 50 GHz - 2 MV/m, while in 100 to 500 GHz - 4 MV/m. Since the peak EM-field is proportional to the square root of power divided by area, the peak EM-field increases with the frequency proportionally to the square root of $2/\lambda$. The increase that follows the curve of the air and vacuum breakdown EM-field of Fig. 1 permits generation of peak EM-fields very close to breakdown level in the entire electromagnetic spectrum. In the 1 to 5 GHz band, the peak E-field from a single broadband generator and TEM-horn antenna exceeds by at least one order of magnitude the 70 kV/m peak EM-field achieved in current narrowband and broadband HPM systems [12]. If instead of one antenna and one generator one antenna with two or four generators is used, the radiated power will increase up to four times [13, 14].

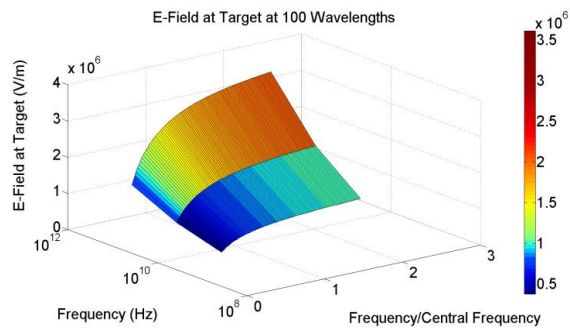

Fig. 9. TEM-horn peak E-field at 100 $\lambda$ from the antenna in function of central and normalized frequency - calculated for antenna mouth width of 2.25 $\lambda$ at the central frequency.

## IV. BROADBAND HPM ARRAYS

### A. Introduction

Obtaining the MV/m EM-field at microwave frequencies is only possible if sub-nanosecond pulses formed by individual generators are used (See Fig. 1). In the narrowband systems, delay in reaching the full peak power of the generated pulses extends the pulse duration into nanoseconds resulting in the generation of EM-fields an order of magnitude lower than the required MV/m. Furthermore, considering, the unknown frequency response of the target, use of narrowband systems with unpredictable coupling of the EM energy to target and low generated EM-field undermines the effectiveness of the narrowband HPM. Therefore, only broadband excitation using multiple sub-nanosecond pulses allows the balanced delivery of energy and power to target. However, considering the kHz repetition rate of a single broadband generator, use of an antenna array with many generators permits realization of multiple sub-nanosecond pulses. Multiple sub-nanosecond pulses, obtained by separate triggering of individual generators, allows varying total pulse duration and patterns, i.e. simultaneous, sequential or combinations that allow one to balance the delivery of energy and power to target. The balancing in turn delivers two basic damage mechanisms – high energy density and high EM-field – otherwise not achievable in the narrowband systems. Generation of EM-fields in the MV/m range enables the deployment of energy capable of damaging a small area of semiconductor devices in picoseconds without burning the entire device and therefore reduces demand for the energy required to destroy the entire device [19]. Additionally, the use of short broadband EM-field pulses in the GV/m range at microwave frequencies allows the manipulation of a chemical reaction at the molecular level. In some cases, short pulse broadband excitation can induce a narrowband resonating target response that extends the excitation effects for a period proportional to the oscillation quality factor. For example, if the oscillation quality factor for cable coupling is in order of 5 to 10, the effect of single pulse excitation will be prolonged up to 10 times. Application of additional short broadband pulses will further prolong the effect of excitation. In summary, the broadband arrays enable one to achieve high EM-field and energy density, adjust pulse spectral content, amplitude, duration, inter-pulse spacing and the sequence of generated pulses. These factors combine to result in the greatest impact on target by tailoring generated EM fields to specific targets. Ref. 4 introduced broadband arrays in flat, concave and convex configurations (See Fig. 10). However, use of flat faced arrays precludes advantageous illumination when each broadband antenna fires pulses at different times and directs them to different areas. The concave configuration allows the illumination of a single target while convex configurations permits uniform illumination of many targets or large objects with each antenna radiating into a different section of the object. For reference see, MIL-STD-464C where 2 x 2 m adjacent areas of a single target are illuminated separately [7].

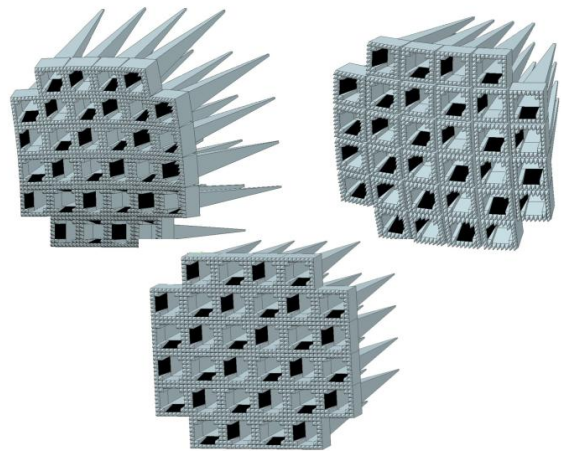

Fig. 10. Broadband TEM-horns (see Fig. 4) assembled into 32 antenna double-polarization concave, convex and flat arrays.

Generation of MV/m EM-field while using broadband arrays depends on the synchronized triggering of individual generators and beam collimation. In this paper, the performance analysis of broadband arrays uses a double exponential pulse (peak amplitude, rise time and duration) matching the bandwidth of the array. Although, the firing of each generator can include time gaps, changing the pulse sequence and varying the spectral content of pulses, this paper addresses only two simplified cases: In the first, the individual generators fire simultaneously, while in second instance the generators fire sequentially with no gaps between pulses. Time jitter in triggering of individual generators affects the EM-field only during simultaneous firing of generators. Figs. 11 and 12 display the results of the LTspiceIV jitter calculation for a 32-antenna array operating in the 1 to 5 GHz band. To allow accurate assessment of jitter on the generated EM-field the calculation assumes no-jitter and the worst case of uniformly distributed jitter of 100, 130 and 160 ps maximum through 32 antennas. The effect of maximum 160 ps jitter on the generated peak EM-field was found to be negligible since the peak EM-field does not decrease by more than 1.5 dB and the rise time of the generated EM-field increases only from 100 to 120 ps (See Fig. 11). Maximum jitter of 160 ps resulted in 10 dB decrease in the amplitude of the spectral component of the EM-field at the maximum frequency of 5 GHz (See Fig. 12). Analysis suggests that achieving 160 ps jitter in the 1 to 5 GHz band requires a coaxial trigger arrangement while at higher frequencies use of ps and fs laser triggering would be necessary.

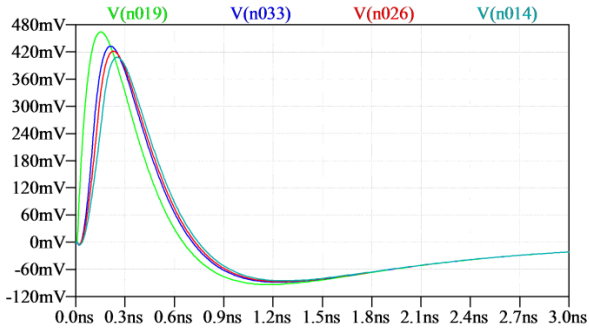

Fig. 11. E-field generated by 32 antennas array operating in the 1 to 5 GHz band. Double exponential pulses of 100 ps rise time and 1 ns duration from 32 generators are matching the bandwidth of the array and triggered with total uniform jitter distribution of: 0 ps - trace n019, 100 ps - trace n033, 130 ps - trace n026 and 160 ps - trace n014.

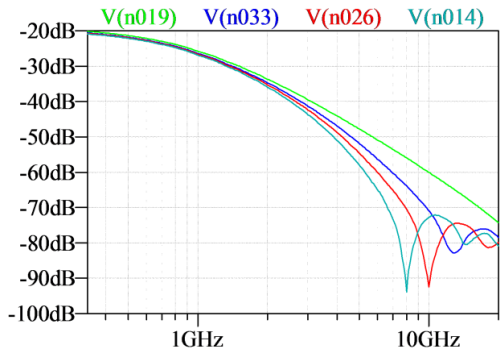

Fig. 12. FFT calculation of spectral content of E-field of Fig. 11.

### B. Broadband HPM TEM-Horn Array

For low power Continuous Wave (CW) and broadband MV/m, HPM testing of immunity at GHz frequencies a concave array consisting of numerous TEM-horns mounted in alternate EM- and H-field polarization was developed [4]. This paper uses similar arrays for direct illumination of targets at short or intermediate distances while long distance illumination was achieved using the array in a Cassegrain antenna. At short-range distance from the array, there is no beam collimation nor there is at the secondary reflector of the Cassegrain antenna. As such, the EM-field at short distances from the array consists of the conflated power from all beams divided by the illuminated area determined through the expanding diameter of individual antenna beams at a specified distance. However, for intermediate target distances where the concave array with simultaneously triggered generators is used, the EM-field relies on beam collimation that reduces the illuminated area proportionally to the number of antennas in the array. Therefore, for the short-range illumination the EM-field decreases with distance from the array, while for intermediate-ranges the EM-field remains constant for an extended distance from the array dictated by the number of antennas. Since the peak EM-field depends on the number of simultaneously triggered antennas, their position in the array and the distance from the array, an accurate calculation of the EM-field requires numerical analysis. Fig. 13 and 14 show examples of EM-field for collimated and non-collimated beams that at intermediate distances, depending on the triggered pulses sequence, vary by an order of magnitude. Examples of EM-field variation in function of distance at intermediate-range use a 32-antenna array operating in 1 to 5 GHz band, with each antenna having the mouth width of 2.25 λ with a central frequency of 2 GHz and the main beam angle of 20 deg. At short-range distances from the array, i.e. 15 m (100 λ), each antenna illuminates the same 5 m (27 λ) diameter target. Assuming 50% antenna efficiency, with each antenna in the array powered by individually triggered generators with 100 ps rise time and 1 ns pulse duration, each antenna radiates 80 GW (Chapter II). Therefore, for a 5 m diameter target, the peak EM-field amplitude for simultaneous triggering of all generators is 4.5 MV/m for 1 ns total pulse duration, and for sequential it is 0.8 MV/m for 32 ns total pulse duration. The peak EM-field of 4.5 MV/m is in the range of air breakdown limit of 6 MV/m. The peak EM-field achieved at short ranges is two orders of magnitude higher than that generated by current narrowband and broadband HPM systems [12]. At intermediate distances from the array, i.e. 75 m (500 λ), simultaneous triggering of the generators led to the successful illumination of the 5 m diameter target, resulting in 4.5 MV/m peak EM-field amplitude at target for 1 ns total pulse duration – matching the results at 15 m distance. However, at intermediate distances and using sequential triggering (no collimation), the peak EM-field decreases for 32 ns total pulse duration to 170 kV/m. In summary, the 32-antenna array with 2.5 m diameter and 1 m antennas length, operating in 1 to 5 GHz allows the achievement of peak EM-fields close to air and vacuum breakdown level. For higher EM-fields, larger testing volume and longer excitation time the number of antennas in the array should be increased (See Fig. 13 and 14.)

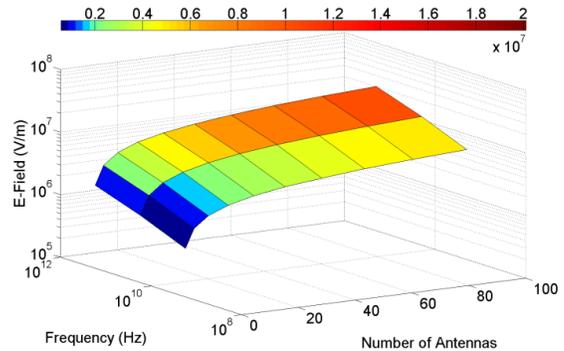

Fig. 13. Peak E-field at a distance of 100 λ from TEM-horn array in function of frequency and number of antennas - no collimated beam formed. Antennas efficiency 50 %, at central frequency the mouth width is 2.25 λ and bandwidth $f_{max}/f_{min}=5$.

Although the presented example addressed the frequency in relatively limited 1 to 5 GHz ranges, extending the graph of Figs. 13 and 14 demonstrates the possibility of designing antenna arrays that can operate up to 500 GHz.

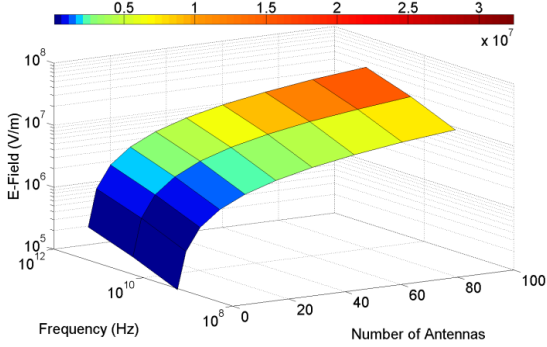

Fig. 14. Peak E-field at a distance of 500 λ from TEM-horn array in function of frequency and number of antennas - collimated beam formed. Antennas efficiency 50 %, at central frequency the mouth width is 2.25 λ and bandwidth $f_{max}/f_{min}=5$.

### C. Broadband HPM Cassegrain Antenna

Cassegrain antennas, on-the axis and offset, deliver the small beam divergence and large angular amplification required to achieve a small skew angle making them ideal for broadband HPM illumination of distant targets. We analyzed the somewhat simpler on-the-axis Cassegrain antenna (Fig. 15) with a feed antenna array and both primary and secondary reflectors located on the same axis. The selected geometry results in an aperture blocking, which was minimized by making the primary reflector substantially larger than the secondary reflector. The beam dispersion and skew are optimized to achieve ± 3 dB E-field variations at the target. Since there is no beam skewing from centrally positioned antennas in the array, for an optimized design of Cassegrain antenna, the number of TEM-horns in the array has to be limited otherwise the angular amplification of Cassegrain antenna needs to be increased. Considering the beam dispersion and skew, the number of antennas $N$ in the array is equal to:

$$N = \frac{\pi}{4}\left(\frac{k_a}{w_\lambda}\right)^2 D_\lambda^2 \quad (2)$$

Equation (2) shows $k_a = \frac{d_0}{D}$, $D$ as diameter of Cassegrain primary reflector in meters, $D_\lambda$ as diameter expressed in wavelength and $\lambda$ is the wavelength at the central frequency of the band (See Equation 1). Diameter of the antenna array is $d_0$ and $w$ is the mouth width of each TEM-horn antenna in the array. The ratio between diameter of the antenna array and the Cassegrain primary reflector is $k_a$, with $w_\lambda$ being TEM-horn antenna mouth width expressed in wavelength. Although, $k_a$ and $w_\lambda$ could vary ($k_a \approx 0.3\ to\ 0.01, w_\lambda \approx 1\ to\ 5$), for an optimized design $k_a=.25$ and $w_\lambda=2.25$ are used, and an optimal number of antennas in the array is:

$$N_{opt} = \frac{\pi}{350} D_\lambda^2 \quad (3)$$

For the optimal number of antennas in the array, the beam farthest from the Cassegrain antenna axis is pointing at the secondary reflector at an angle $\varphi_a =0.1745$ rad. Considering angular amplification of the Cassegrain antenna $m_a \approx 10$, the optimal skew angle is $\beta_t= \varphi_a/m_a=0.01745$ rad. Therefore, to maintain the EM-field uniformity at the target, the main beam divergence and skew angles are set to be identical and consequently the primary reflector diameter $D_\lambda$ and the main beam diameter at the target $D_{t\lambda}$ located at distance $R_\lambda$ are:

$$D_\lambda \leq \frac{2}{\sqrt{\pi}*\beta_t} \quad (4a)$$
$$D_{t\lambda} = \beta_t * R_\lambda \quad (4b)$$

Equation 4(a) and 4(b) show a relationship between primary reflector diameter $D_\lambda$ and the skew angle $\beta_t$. For collimated beam extending into the target $D_\lambda=D_{t\lambda}$ the primary reflector diameter $D_\lambda$ and target distance $R_\lambda$ are correlated. For an antenna having diameter of primary reflector, $D_\lambda = \beta_t * R_\lambda$, the beam skew does not affect the performance of the antenna since the main beam and the skew diverge at the same rate.

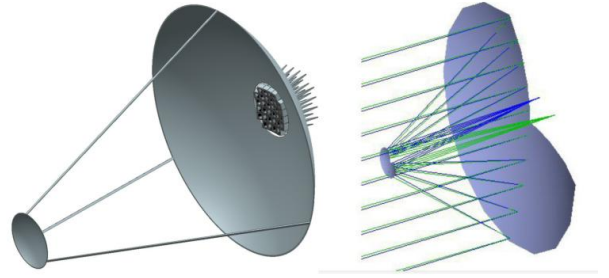

Fig. 15. Physical and "OSLO" simulation view of high power broadband Cassegrain antenna.

Maximum peak EM-field for an optimally designed Cassegrain antenna having primary reflector and target diameter equal to 60 λ, illuminated by 32 antenna array operating in the 1 to 500 GHz band are shown below. The 60 λ diameter was chosen to correspond with 9 m diameter antennas in current narrowband HPM facilities [20], operating in the 1 to 5 GHz band. For other frequencies, the antenna diameter and target distance were frequency scaled resulting in the following calculated peak EM-fields:

- In 1 to 5 GHz band at distance R≤ 600 m, 9 m diameter antenna delivers to 9 m target E-field of 3 MV/m for 1 ns total pulse duration, for all simultaneously triggered generators and 0.5 MV/m for 34 ns for sequentially triggered.
- In 10 to 50 GHz band at distance R≤ 60 m, 1 m diameter antenna delivers to 1 m target EM-field of 5 MV/m for 100 ps total pulse duration, for all simultaneously triggered generators and 1 MV/m for 4.2 ns for sequentially triggered.
- In 100 to 500 GHz band at distance R≤ 6 m, 10 cm diameter antenna delivers to 10 cm target EM-field of 9 MV/m for 10 ps total pulse duration, for all simultaneously triggered generators and 1.6 MV/m for 420 ps for sequentially triggered.

For 1 to 5 GHz band, the achieved 3 MV/m peak EM-field is almost two orders of magnitude higher than 69 kV/m, which is currently accepted as the maximum narrowband HPM threat [7]. Therefore, if current 69 kV/m limit of HPM threat applies, the findings of this paper indicate that it is possible to generate such fields on targets located 25 km away.

### D. Broadband HPM Cassegrain Antenna with Barlow Lens

If the collimated beam range needs to be extended to illuminate a target visible at angles smaller than the optimal skew angle $\beta_t$= 0.01745 rad. it becomes necessary to reduce the skew. A Cassegrain antenna with an added Barlow lens (Fig. 16) makes it possible to increase the angular amplification and reduce the skew angle.

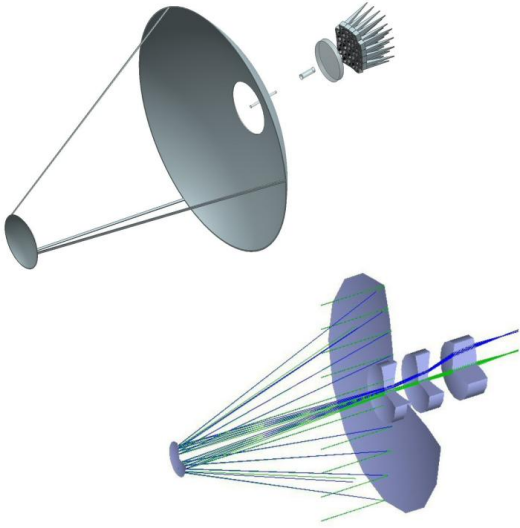

Fig. 16. Physical and "OSLO" simulation view of high power broadband Cassegrain antenna with Barlow lens.

Considering that, the maximum amplification of Cassegrain antenna is approximately 10, the addition of Barlow lenses using a multiplication factor of 60 increases the amplification to 600 without causing optical instability. However, should such amplification be used in high power systems this may result in surface breakdown at the third Barlow lens. To avoid breakdown in high power systems the angular amplification should be limited to not more than 10. Equation 5 shows the calculation of Barlow lens amplification $m_b$ required to extend the collimated beam to a distance $R$:

$$m_b = \frac{\varphi_a}{\beta * m_a} \quad (5)$$
$$\beta = \frac{D_t}{R}$$
$$\varphi_a = \frac{d_0}{F(1 - \frac{d_r}{D})}$$
$$m_a = \frac{L + F\left[1 - \frac{d_r}{D}\left(1 - \frac{\Delta B}{F}\right)\right]}{F \frac{d_r}{D}\left(1 - \frac{\Delta B}{F}\right)}$$
$$\Delta B = 2F - \sqrt{(2F)^2 - \left(\frac{D}{2}\right)^2}$$
$$D_t = \frac{2}{\sqrt{\pi}} \frac{R}{D_\lambda}$$

In Equation 5, $\varphi_a$ is the angle where the antenna furthest from the axis in the array is pointing at the centre of the secondary reflector. The distance between the focal point (located inside the farthest from the axis TEM-horn) and the centre of the primary reflector is $L$, $\beta$ is the angle defined as a target diameter divided by target distance from the centre of the primary reflector and $d_0$ is diameter of the antenna array. The diameter of the secondary reflector is $d_r$ and $F$ is the focal distance from the primary reflector. At non-diverging distances, the calculated peak EM-field levels while using Cassegrain antenna with 100 antennas in the array and Barlow lens with angular amplification of two are:

– In 1 to 5 GHz band at distance R≤ 1.5 km, 15 m diameter Cassegrain antenna supplies to 15 m target E-field of 3 MV/m for 1 ns total pulse duration for all simultaneously triggered generators and 300 kV/m for 100 ns for sequentially triggered.
– In 10 to 50 GHz band at distance R≤ 150 m, 1.5 m diameter Cassegrain antenna supplies to target of 1.5 m E-field of 6 MV/m for 100 ps total pulse duration for all simultaneously triggered generators and 600 kV/m for 10 ns for sequentially triggered.
– In 100 to 500 GHz band at distance R≤ 15 m, 15 cm diameter Cassegrain antenna supplies to target of 15 cm E-field of 10 MV/m for 10 ps total pulse duration for all simultaneously triggered generators and 1 MV/m for 1 ns for sequentially triggered.

In the 1 to 5 GHz range, the peak EM-field of 3 MV/m at 15 km distance decreases to 300 kV/m, and at 65 km to 69 kV/m level defined by a current HPM standard. The main benefit of using a Barlow lens is the ability to extend the target range, and to increase diameter of the target and number of antennas in the array. The higher number of antennas in the array makes it possible to increase the number of pulses generated resulting in delivery of higher energy to the target.

## V. COLLIMATED BROADBAND HPM SPHERICAL ARRAYS FOR FUSION PLASMA STUDIES

The frequencies of fusion plasma are in the 300 GHz range. Currently, fusion plasma generation utilizes microscopic excitation at frequencies below the indicated range, laser excitation at frequencies above and low power microwaves within the range. Two of the indicated methods do not allow plasma excitation at resonance frequencies and the third method does not provide sufficient power density required for fusion. However, collimation of 192 beams generated from on-the-axis Cassegrain antennas at a single point and distance of four focusing lens diameters (Fig. 17), allows illumination of target using ps pulses at ionizing GV/m peak E-field in 1 to 500 GHz band that includes the plasma fusion frequencies.

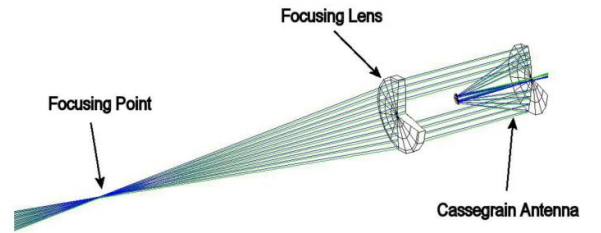

Fig. 17. "OSLO" simulation view of Cassegrain antenna radiated parallel beam collimated into a focal point.

Furthermore, broadband HPM excitation using multiple independently triggered pulses permits variance in time the energy and peak power delivered to target. By carefully selecting the spectral content, amplitude, duration and inter-pulse spacing of the pulses the excitation of plasma resonances al-lows maximizing the plasma response. Fig. 18 shows a comparison of vacuum and air breakdown EM-field with peak EM-field radiated by 192 Cassegrain antennas. For simultaneous triggering, the generated peak EM-field is in the GV/m range and exceeds by an order of magnitude the limits of the breakdowns field in the air (molecular breakdown), and vacuum (plasma generation) established in the 1992 CT concept. For sequential triggering, the peak EM-field is below both breakdown limits. By varying the sequence of pulse excitation, the EM-field at the target could be either below or above the breakdown level allowing optimized excitation. Plasma excitation in the entire 1 to 500 GHz range allows one to choose the plasma resonances region depending on plasma density and temperature. Furthermore, decreasing pulse duration (higher frequency of excitation), increases the plasma breakdown voltage which in turn allows generation of higher peak EM-field and in greater energy deposition into plasma. The energy level in broadband pulse excitation increases when firing the generators repeatedly at the speed related to the charging time of Marx generators. Additionally, in the 100 to 500 GHz band, the diameter of illumination is in the range of 1 to 10 mm corresponding with the size of fusion targets (See Table 1). The short pulse broadband excitation induces a narrowband resonating plasma response and prolongs the effects of excitation for a period proportional to the oscillation quality factor with no need for supplying additional power during the plasma oscillation. Since it is not necessary to supply additional power the fusion energy balance improves and reduces the requirement for cooling creating a new paradigm for the molecular and atomic research leading to fusion. Considering that the magnetic confinement field required for fusion is in the range of 10 T, it is possible to place a small 70 cm diameter spherical array in a currently available 10 T MRI magnet.

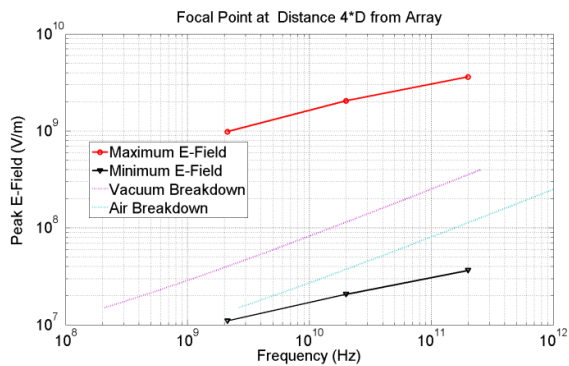

Fig. 18. Comparison of vacuum and air breakdown with peak EM-field radiated by 192 Cassegrain antennas of Fig. 15 to a single focal point in function of frequency equivalent to the pulse duration $f=1/T$. Single focal point is 4 beam diameters away from spherically arranged lenses.

## VI. Conclusions

The capability of the broadband HPM systems presented here substantially exceeds capability of current narrowband and broadband HPM systems, while antenna dimensions, power supply and system weight is either smaller or similar. With appropriate design and development of support infrastructure, it is now possible to deploy HPM-technology in both military and non-military applications. The ultimate broadband TW and PW High Power Microwave systems operating in the 1 to 500 GHz band result in unprecedented:

- Delivery of the ultimate radiated electromagnetic threat peak EM-field that corresponds to air or vacuum breakdown levels.
- TW immunity test facilities that permit hardening and testing of large systems at the Ultimate Threat levels assuring pulse spectral content matching the most susceptible spectral region of the target.
- Broadband HPM Cassegrain antenna arrays and Cassegrain antenna arrays with Barlow lens, that in 1 to 5 GHz band at TW power permits irradiation of target in the distance range of 10's km with a pulse duration comparable to duration in narrowband systems and exceeding pulse duration of current broadband systems.
- Broadband HPM spherical arrays for collimated plasma excitation using ps pulses at GV/m EM-field levels at fusion plasma frequencies in the range of 300 GHz that by varying in time the energy and peak power, the pulse spectral content, duration and inter-pulse spacing maximizes the plasma response.

TABLE I
Spherical Arrays for Collimated Plasma Studies

| Frequency Range (GHz) | 1 to 5 | 10 to 50 | 100 to 500 |
|---|---|---|---|
| Peak Power per Generator (GW) | 160 | 6.4 | 0.2 |
| Radiated Peak Power per Antenna (GW) | 80 | 3.2 | 0.1 |
| Diameter of Cassegrain Antenna (m) | 8 | 1 | 0.1 |
| Number of Antennas in Cassegrain Ant. | 42 | 52 | 52 |
| Peak Power per Cassegrain Ant. (GW) | 3400 | 166 | 5.2 |
| Number of Cassegrain Antennas | 192 | 192 | 192 |
| Total Number of Antennas | 8064 | 9984 | 9984 |
| Max. Power at Target (TW) | 645 | 32 | 1 |
| Target Diameter (cm) | 50 | 6 | 0.6 |
| Max. Power Density at Target (PW/m+2) | 3.28 | 11.3 | 35.4 |
| Max. E-field at Target (GV/m) | 1 | 2.1 | 3.65 |
| Min. E-field at Target (MV/m) | 12.4 | 21 | 36.5 |
| Min. Time for Max. E-field (ps/discharge) | 1000 | 100 | 10 |
| Max. Time for Min. E-field (ns/discharge) | 8000 | 3200 | 100 |
| Max. Number of Discharges (1/s) | 2*10+3 | 2*10+4 | 2*10+5 |
| Energy on Target (kJ/discharge) | 645 | 3.2 | 0.01 |
| Total Energy on Target (MJ) | 1300 | 64 | 2 |
| Energy Density on Target (J/cm+2/discharge) | 328 | 113 | 35 |
| Total Energy Density on Target (kJ/cm+2) | 657 | 2264 | 7100 |
| Time to Reach 10+4 kJ/cm+2 (s) | 15 | 5 | 1.4 |
| Weight of Target (g) | 2*10+5 | 3400 | 0.33 |
| Radius of Focusing Facility (m) | 32 | 3.5 | 0.35 |

## Appendix

### Current HPM Threats and Test Facilities

The compilation of current threats in accordance with Military standard MIL-STD-464C [7] shown in Fig. 19, indicates that the most severe narrowband intra-system EM environments are limited to peak EM-field levels of 28 kV/m$_{RMS}$ in the 2.7 to 3.6 GHz band. For intentional threats, the MIL-STD-464C stipulates that at a distance of 1 km and frequencies from 8.5 to 11 GHz, the narrowband EM-field peak level is 69 kV/m. However, due to increasing concerns regarding the values specified by the standard, the MIL-STD-464C cautions that the narrowband intentional threat specifications "should be verified prior to implementation."

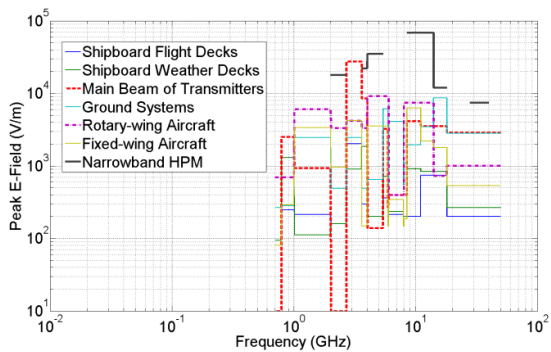

Fig. 19. Maximum HPM EM-field levels - MIL-STD 464C.

Current narrowband HPM facilities are designed to operate in the frequency range from 0.7 to 3 GHz at peak power of 0.4 GW [20]. The maximum radiating antenna size is limited to 7 x 7 m resulting in 32 dBm antenna gain. Max. EM-field level, being in the range of 60 to 75 kV/m, corresponds with the value specified by MIL-STD-464C. The generated peak power and the maximum EM-field required limit the dimensions of the test area to 7 x 7 m and the pulse duration to 500 ns. The only available data for Orion GB/USA test facility (Orion), indicates that to sustain the pulse repetition frequency of 100 Hz at GW peak power, 500 kW power supply is required. Such power requirement restricts the mobile applications of narrowband facility. To cover the entire 1 to 3 GHz frequency range, the narrowband coupling characteristic of the tested system (Q=5 to 10), demands difficult and time consuming tuning of high power generators. As such, in case of need for quick verification of system immunity, use of narrowband testing systems results in omitting the effect of narrowband coupling [19]. To address the issues related to the use of narrowband HPM test facilities: portability, shortening the test time and reducing the cost of the facilities, very early on portable, broadband, GW NEMP/HPM facilities were developed in Canada and other countries [ 4, 13, 14, 17, 18, 21 ].

**Andrew S. Podgorski** (M'76–SM'86–F'08) holds M.A.Sc. and Ph.D. degrees

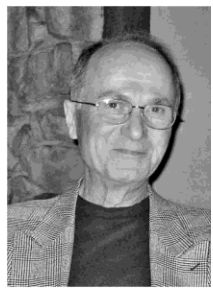

in electrical engineering from the University of Waterloo, Canada in 1975 and 1980. During his study, he was a Research Assistant with the Microwave Laboratory engaged in numerical modeling of dielectric filters, distributed THz semiconductor devices and Very Large Array radio astronomy observatory in Socorro, NM.

Prior to that, from 1969 to 1973, he was a researcher at Siemens Microwave Laboratory, Vienna, Austria, engaged in design of microwave communication systems for many of the European satellites. From 1980 to 1995, he was the Head of the EM Protection Group at the National Research Council of Canada and an Adjunct Professor at the University of Ottawa, Canada. His research expanded knowledge of lightning from µs to ns, ESD from ns to ps, HPM from ns to sub-ps, and resulted in 3D numerical modeling of LEMP, ESD and HPM interactions. From 2004 to 2007, he was the US National Research Council Resident Research Associate at Air Force Research Laboratory in Albuquerque, NM, conducting studies of broadband THz HPM antennas. Since 1995, he has been President of ASR Technologies Inc., a private company conducting research in broadband sub-picoseconds high power electromagnetics that allowed development of broadband multi-decade high power antennas and generators, automated broadband LEMP, EMI/EMC, NEMP and HPM test and measurement facilities. He has authored numerous publications and classified and non-classified reports and he holds many patents in the area of broadband electromagnetics. His current scientific interest concentrates on plasma and particle physics in application to high power sub-picoseconds electromagnetic phenomena.

Dr. Podgorski has served on the Board of Directors for the Electromagnetic Compatibility Society of IEEE serving as the Chair of the Technical Activities Committee, Treasurer and a "Distinguished Lecturer." He is a Life Fellow of the IEEE, Honorary Life Member of the EMC Society and he is listed in a Canadian "Who's Who."